\documentclass[12pt]{amsart}
\usepackage{amssymb, latexsym, amsmath, amsfonts, tikz, tikz-cd}
\usepackage{hyperref}

\theoremstyle{definition}

\theoremstyle{remark}

\numberwithin{equation}{section}
\theoremstyle{remark}

\begin{document}
\title[Game-theoretic approach to behavioural interaction in disease dynamics]{An evolutionary game theory approach to modeling behavioural interaction in disclosing infection begins with an outbreak: COVID-19 as an example}

\author{Pranav Verma, Viney Kumar, Samit Bhattacharyya}

\address{PV: Student, Department of Mathematics, Shiv Nadar University, Delhi - NCR 201314, India }
\email{pv638@snu.edu.in}

\address{VK: Research Scholar, Department of Mathematics, Shiv Nadar University, Delhi - NCR 201314, India}
\email{vk981@snu.edu.in}

\address{SB: Professor, Department of Mathematics, Shiv Nadar University, Delhi - NCR 201314, India}
\email{samit.b@snu.edu.in}

\begin{abstract}
The global impact of the COVID-19 pandemic on the livelihoods of people worldwide prompted the implementation of a range of preventive measures at local, national, and international levels. Early in the outbreak, before the vaccine became accessible, voluntary quarantine and social isolation emerged as crucial strategies to curb the spread of infection. In this research, we present a game-theoretic model to elucidate the voluntary disclosure of exposure to infected individuals within communities. By employing a fractional derivative approach to illustrate disease propagation within the compartmental model, we determine the minimum level of voluntary disclosure required to disrupt the chain of transmission and allow the epidemic to fade. Our findings suggest that higher transmission rates and increased perceived severity of infection change the externality of disclosing infected exposure, thereby contributing to a rise in the proportion of individuals opting for quarantine and reducing disease incidence. We estimate behavioural parameters and transmission rates by fitting the model to hospitalized cases in Chile, South America. Results from our paper underscore the potential for public health authorities to influence and regulate voluntary disclosure of infection during emerging outbreaks through effective risk communication, emphasizing the severity of the disease and providing accurate information about hospital capacity to the public. 
\end{abstract}



\maketitle
\section{Introduction}
\begin{figure}
\begin{center}
\includegraphics[scale=0.4]{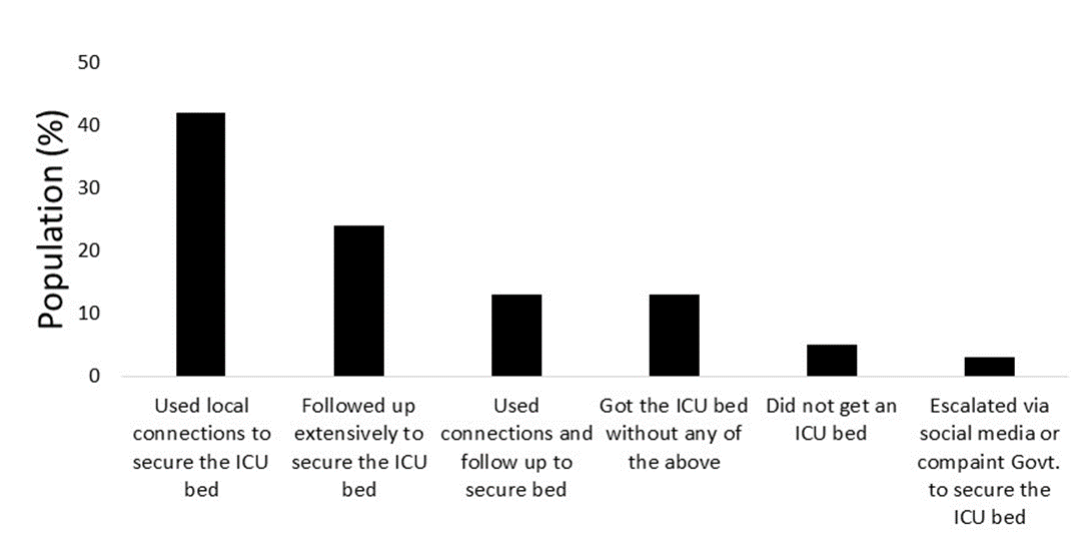} 
\caption{Impact of coronavirus (COVID-19) on securing ICU beds in hospitals across India as of April 2021 (Data source: \href{https://bit.ly/4hJ0xT4}{Link to the source})}
\label{FIG:0}
\end{center}
\end{figure}

The COVID-19 {}{disease caused by the SARS-CoV-2 virus}, which was first detected in China in December 2019  \cite{kraemer2020effect}, caused an unrivalled worldwide outbreak of infections, with 95 million reported cases globally \cite{gebru2021global,lee2021impact}. World Health Organization (WHO) officials moved quickly to declare the COVID-19 pandemic a global health emergency in reaction to the worsening situation \cite{jee2020international,kaur2020covid}. Mandatory lockdowns, quarantines, self-isolation, and the wearing of masks were among the measures put in place by authorities at the beginning of the outbreak to restrict the spread of the virus \cite{cevik2021support,pentecost2022effects}. Clinical practices such as rigorous testing, quarantine measures, and close monitoring were also initiated \cite{liu2020modelling,yang2022spread}. A critical shortage in hospital bed availability and a deficit in essential medical resources like oxygen and medication marked a significant crisis \cite{kapoor2023impact,tirupakuzhi2021challenges}. By the end of September 2020, India had reported over $63 \times 10^5$ confirmed COVID-19 cases, exhibiting a range of symptoms from mild upper respiratory tract issues to severe conditions such as acute respiratory distress syndrome and multi-organ failure, necessitating intensive care.\\
Even though a relatively small percentage of patients required critical care services, the surge in cases quickly overwhelmed the healthcare system. At the beginning of the COVID epidemic in India, the Ministry of Health and Family Welfare (MoHFW) suggested that around 2.5\% of patients needed intensive care. However, this might be underestimated due to incomplete reporting in some states \cite{ghosh2020india}. As of June 28, 2020, the MoHFW reported 1,055 dedicated COVID hospitals in India, with 177,529 isolation beds and 78,060 oxygen-supported beds. Additionally, there were 2,400 dedicated COVID Health Centres with 140,099 isolation beds and 51,371 oxygen-supported beds. However, this still left a considerable gap, with approximately 120,000 oxygen-supported beds available. It was estimated that 15\% of patients, translating to about 1.5 million individuals in India, would require mild to moderate infection treatment with oxygen beds. Securing beds posed a significant challenge, as revealed by a Local Circles survey in April 2021 (\href{https://bit.ly/4hJ0xT4}{www.statista.com}). Only 13\% of respondents successfully obtained an ICU bed through the standard procedure, while the majority had to rely on personal connections. The survey indicated difficulties securing COVID-19 ICU beds for family and friends (see Fig \ref{FIG:0}).
In confronting the critical challenge of disease outbreak management, health authorities consistently advocate for voluntary adherence to quarantine measures, stressing the significance of individuals disclosing potential exposure and initiating self-quarantine protocols \cite{zachosova2021management,schiffer2021moral}. {}{In fact, it has been found that disease transmission can be potentially reduced by increasing such intervention policies}. However, the decision to comply with such directives involves intricate strategic deliberations at the individual level, weighing perceived benefits against associated costs, and encompassing economic ramifications and constraints on personal liberties \cite{satapathi2023coupled,saad2023dynamics,premkumar2023impact}. To thoroughly examine the dynamic interaction between social distancing, quarantine compliance, and disease containment, researchers have increasingly turned to game-theoretic frameworks (\cite{funk2010modelling} and references therein), especially in vaccination behaviour \cite{bhattacharyya2011wait, bauch2012evolutionary}, social-distancing \cite{bhattacharyya2019game}, self-medication of antibiotics \cite{malik2022disparity} and {}{Insecticide-treated nets (ITN)} usage in malaria control \cite{laxmi2022evolutionary}. By utilizing game theory concepts within epidemiological models, such as the SEIR model, A. Satapathi et al., \cite{satapathi2023coupled} have investigated how individuals make decisions to safeguard themselves and engage with one another. This integration enables exploring diverse scenarios and behavioural responses to quarantine mandates.

During the COVID-19 outbreak, several distinct computational and mathematical models have been developed to investigate the significance of quarantine as a game-theoretic strategy {}{and its impact on disease transmission as well as hospitalization rates} \cite{saad2023dynamics, martcheva2021effects,alam2022game,khazaei2021disease}. C. N. Ngonghala et al. \cite{martcheva2021effects} introduced two models inspired by the COVID-19 pandemic, incorporating elements of game theory, disease dynamics, human behaviour, and economics. A study by A. Rajeev et al. \cite{premkumar2023impact} delved into an evolutionary game theory model to examine individuals' behavioural patterns and identify stable states. {}{M. Alam} et al. \cite{alam2022game} emphasized the significance of implementing multiple provisions promptly to enhance disease containment efforts using game theory and human behaviour. H. Khazaei et al. \cite{khazaei2021disease} analyzed the SEIR epidemiological model alongside an individual behaviour response model, specifically exploring the dynamics of a game where the public's compliance with social distancing measures is influenced by the state of disease and associated payoffs.\\\\
Using the fractional derivative method, several mathematical models have been constructed to account for modelling memory effects and long-range dependencies, often present in real-world epidemiological data \cite{ullah2022dynamic,yunus2022mathematical,wali2022stability}. To examine the impact of individual vaccination on monitoring COVID-19 transmission, Akeem O. Yunus et al. \cite{yunus2022mathematical} developed {}{and analyzed a} fractional order model. M. Wali et al. \cite{wali2022stability} introduced a novel numerical approach for the COVID-19 epidemic model based on the Atangana-Baleanu fractional order derivative in the Caputo sense to study vaccination efficacy.\\\\
However, no study considers disclosing exposure to infection as an individual choice in the face of an outbreak and limited availability of medical facilities. Understanding the interplay between individual strategies, medical facilities' availability, and its consequences on disease burden is an important challenge for public health policymakers. This study explores the interplay between individual decision-making and epidemiological factors by developing a game-theoretic model utilizing fractional-order derivatives. We have identified the critical level of disclosure necessary for the epidemic to fade out. We simulate and explore how the disease transmission rate and interventions in public health, or the severity of the disease, influence individuals' decisions. According to our results, the population exhibits higher disclosure tendencies due to the combined impact of transmission and disease severity. Our analysis underscores the importance of infection disclosure as a strategic decision and its implications for illness control and the burden of hospitalization.
\section{Model framework}
\begin{figure}
     \centering
    \includegraphics[scale=0.13]{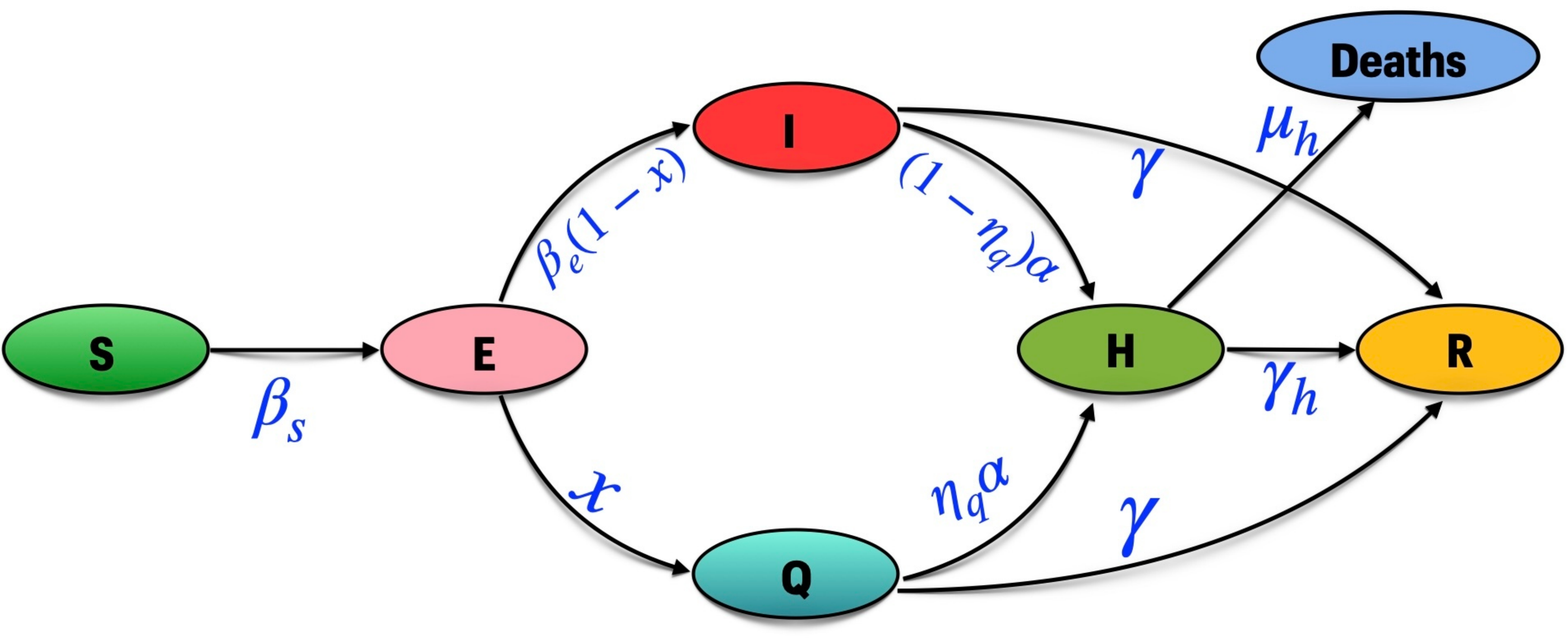}
         \caption{Schematic of the behaviour-prevalence model}
         \label{FIG:1}
\end{figure}
\subsection{Disease model}
We model the human population consisting of six distinct compartments: susceptible (S), exposed (E), infected (I), quarantined (Q), hospitalized (H), and recovered (R). {}{Denoting $S,E,I,Q,H \text{ and $R$}$ as the number of individuals in the respective compartments,} the total population ($N$) is given by $N = S+E+I+Q+H+R$. Figure \ref{FIG:1} displays the schematic of the disease model, and Table \ref{table1:values} describes the parameters used in the model. {}{At the onset of disease spread, a susceptible individual who comes in contact with an infected individual may or may not develop symptoms. We define the exposed state (E) as the subpopulation with a suspected infection. The mean transmission rate of individuals from the $S$ to the $E$ state is denoted by $\beta_s$.} 
{}{Further, we assume that individuals aware of their exposure to infection may choose to disclose their exposure to the infection, voluntarily participate in self-quarantine, and subsequently move to the quarantined (Q) state.} Those who do not opt for quarantine spread the infection to neighbouring individuals. So, we denote this subpopulation as the infected (I) state. In each of the I and Q states, individuals may develop severe symptoms and subsequently {}{need to }be hospitalized. {}{In such a scenario, we assume that the hospitalizations from both states occur at an equal admission rate $\alpha$, coupled with a public health preference factor $\eta_q$ (discussed in the \textit{Game-theory model} description below). Individuals in both these states may also directly recover at a rate $\gamma$ or may recover after being hospitalized at a rate $\gamma_h$.} Additionally, there is a risk of disease-induced death among hospitalized patients, {}{which occurs at a mortality rate $\mu_h$}. Below, we have the epidemiological model equations using the fractional order derivative approach:

Individuals in the exposed or infected class can transmit the infection to susceptible individuals.
\begin{eqnarray}
\label{eq:1}
\nonumber
^{C}_{0}D^{\zeta}_{t}S &=& \frac{-\beta_s(I+E)S}{N}\\
\nonumber
^{C}_{0}D^{\zeta}_{t}E &=& \frac{\beta_s(I+E)S}{N}-xE-\beta_{e}(1-x)E\\
\nonumber
^{C}_{0}D^{\zeta}_{t}Q &=& xE-\eta_{q}\alpha Q - \gamma Q \\
\nonumber
^{C}_{0}D^{\zeta}_{t}I &=& \beta_{e}(1-x)E-(1-\eta_{q})\alpha I-\gamma I\\
^{C}_{0}D^{\zeta}_{t}H &=& (1-\eta_{q})\alpha I+ \eta_{q}\alpha Q-\gamma_{h}H -\mu_hH
\end{eqnarray}
subject to initial condition:
$S(0) = S_0$, $E(0) = E_0$, $I(0) = I_0$, $Q(0) = Q_0$, and $H(0) = H_0$.\\
\noindent We do not explicitly mention the recovery compartment $R$ here in the model equation (\ref{eq:1}), as individuals become completely immune after recovery and the population is closed. $x$ represents the percentage of exposed individuals who choose to disclose the exposure to infection and eventually be quarantined (see game-theoretic model below for a description of $x$).
\begin{table*}
\begin{center}
\caption{Baseline parameters $\&$ variables {}{as described in the game-theoretical model} (\ref{eq:c1})}.\label{table1:values}
\begin{tabular}{cc} \hline
Parameters & Description \\ \hline
$\beta_s$  & Mean disease transmission rate (per day) \\
$\beta_e$  & Mean incubation rate (per day) \\
$\eta_q$  & Preference by public health authorities to provide treatment facility\\
$\alpha$  & Rate of hospital admissions (per day) \\
$\mu_h$  & Mortality rate after hospitalization (per day)\\
$\zeta$ & Order of fractional derivative\\
${\gamma}$  & Recovery rate from infection (per day)  \\
${\gamma_h}$ & Recovery rate of hospitalized patients (per day) \\
$\kappa$ & Sampling rate in social learning (per day) \\
$c_d$ &  Per unit cost of disclosing infection\\
$c_{nd}$ &  Per unit cost of non-disclosing infection\\
$p_S$ &  {}{Perceived probability of developing severe symptoms upon infection}\\
\hline
 Variables & Description \\ \hline
$N$ & Total population  \\
$S(t)$ & {}{Number of }susceptible {}{individuals at time $t$} \\
$E(t)$ & {}{Number of }exposed {}{individuals at time $t$}   \\
$I(t)$ & {}{Number of }infected {}{individuals at time $t$}   \\
$Q(t)$ & {}{Number of }quarantined {}{individuals at time $t$}  \\
$H(t)$ & {}{Number of }hospitalized {}{individuals at time $t$}  \\
$R(t)$ & {}{Number of }recovered {}{individuals at time $t$}  \\
$x(t)$ & Fraction of exposed population {}{opting to disclose exposure at time $t$} \\ \hline
\end{tabular}
\end{center}
\end{table*}
\subsection{Game-theoretical model}
We classify this decision-making framework as a population game, where an individual's payoff is determined by their own behaviour as well as the collective behaviour of the community. The players are individuals exposed ($E(t)$) to infection. {}{i.e., the strategy update takes place after exposure to infection in the context of disease dynamics. It is worth noting that individuals who choose to self-isolate are aware of their potential exposure to infection and voluntarily participate in self-quarantine. The decision-making is not based on symptoms of infection.} Let $x$ ($0 \leq x \leq 1$) represent the fraction of the players who disclose their infection and opt for quarantine. It might be interesting to note that players of the present generation compete not only with players of the previous generation but also with players from past generations who have similar behaviours since the individual choice relies on the current illness prevalence and availability of hospital treatment. We suppose that people imitate other individuals' behaviour, which is more likely for strategic decision-making in various social engagements. In particular, they adopt strategies from other members with a likelihood proportional to the projected pay-off increase if the sampled individual's pay-off is greater \cite{huang2022game,bhattacharyya2010game}. Individuals are presumed to choose their strategy based on the perceived benefits of disclosing the infection and quarantine.\\

In game theory, the players are assumed to maximize their payoff. We consider two strategies: \textit{disclosing infection and opting for quarantine}, and \textit{non-disclosing infection}. The perceived payoff of adopting a disclosure strategy is:
\begin{align}
    \label{eq:G1}
    p_d &= -c_d (1-\eta_q)
\end{align}
where $c_d$ is the per-unit perceived cost of disclosing infection and isolation. We have also assumed that public health authorities prefer providing treatment facilities to individuals, depending on the disclosure of infection. {}{By preferences, we mean incentives in the form of assured medical care (hospital beds and constant health monitoring), monetary benefits and helplines for quarantined individuals who develop severe symptoms} \cite{mitigateindia}. To facilitate this, many centralized isolation and quarantine facilities exist {}{were set up}for patient observation and treatment \cite{wang2021facilities,zhu2021effects}. {}{Such incentives became especially important in the early phases of COVID-19 in middle-income countries like India to promote quarantine as a reliable method of disease containment and reducing the stigma associated with it} \cite{mitigateindia}. {}{Such centralized facilities also play a major role in closely tracking the disease spread, thereby contributing to a more effective implementation of providing timely medical care to those who need it. As a result, these incentives that come along with quarantine become a lucrative choice for the exposed individuals, provided they participate by disclosing their exposure. Hence, we introduce} the parameter $\eta_q$ to {}{encompass all such preferences given} by public health authorities. {}{The fact that the payoff of disclosing} $p_d$ {}{is directly proportional to} $\eta_q$ {}{in} (\ref{eq:G1}) {}{is based on the assumption that greater availability of such public health preferences yields a larger payoff as compared otherwise. Our main intention is to analyze this factor's impact on the players' decision strategies during the disease outbreak. To do so, we simulate our model under both scenarios: when such preferences are available in varying magnitudes (given by} (\ref{eq:G1})) {}{and when they are not} ($\eta_q$ = 0).\\
The perceived payoff of adopting a non-disclosure strategy is:
\begin{align}
    \label{eq:G2}
    p_{nd} &= -c_{nd} p_s f(H)
\end{align}
where $c_{nd}$, $p_s$ are the perceived per unit cost of non-disclosing infection and the probability of developing severe infection, respectively. Individuals’ perceptions of this factor are influenced by the community's current hospital admissions $H(t)$. $f(H)$ is a function indicating that costs increase as $H$ increases:
\begin{align}
    \label{eq:G3}
    f(H) &= \frac{H}{A+H}
\end{align}
where $A$ is the half-saturation coefficient. 
The cost of non-disclosing of infection increases as $H$ increases. So, the expected payoff gain upon changing the non-disclosing strategy to disclosing is:
\begin{align}
    \label{eq:G4}
    \Delta E &= p_d - p_{nd} = -c_d (1-\eta_q)+c_{nd} p_s f(H)
\end{align}
All individuals are payoff maximizers who randomly sample other population members, and {}{adopt that} strategy {}{which} provides them with the highest payoff. Thus, the time evolution of the frequencies of the disclosure strategy for an imitation game is as follows:
\begin{align}
    \label{eq:G5}
    ^{C}_{0}D^{\zeta}_{t}x &= \kappa x(1-x)(-c_d(1-\eta_q)+c_{nd}p_s f(H))
\end{align}
Here, $\kappa$ is the sampling rate in social learning. This is similar to the replicator equation in evolutionary game theory \cite{hofbauer1998evolutionary,bauch2005imitation}.
\subsection{Coupled disease game-theoretic model}
Combining the game-theoretic model (equation \ref{eq:G5}) and the epidemiological model (equation \ref{eq:1}), we have:
\begin{eqnarray}
^{C}_{0}D^{\zeta}_{t}S &=& \frac{-\beta_s(I+E)S}{N} \nonumber\\
^{C}_{0}D^{\zeta}_{t}E &=& \frac{\beta_s(I+E)S}{N}-xE\nonumber-\beta_{e}(1-x)E \nonumber\\
^{C}_{0}D^{\zeta}_{t}Q &=& xE-\eta_{q}\alpha Q - \gamma Q \nonumber\\
^{C}_{0}D^{\zeta}_{t}I &=& \beta_{e}(1-x)E-(1-\eta_{q})\alpha I-\gamma I \nonumber\\
^{C}_{0}D^{\zeta}_{t}H &=& (1-\eta_{q})\alpha I+ \eta_{q}\alpha Q-\gamma_{h}H -\mu_hH \nonumber\\
^{C}_{0}D^{\zeta}_{t}x &=& \kappa x (1-x) (-c_d (1-\eta_q)+c_{nd}p_s f(H))\label{eq:c1}
\end{eqnarray}
subject to the initial condition:
\begin{align*}
    S(0) = S_0, E(0) = E_0, I(0) = I_0, Q(0) = Q_0,\\ H(0) = H_0, x(0) = x_0
\end{align*}
We evaluate and computationally simulate this model (equation \ref{eq:c1}) to learn how changing behavioural and epidemiological characteristics affect the dynamics of disease prevalence.

\section{Results}
\subsection{Reproduction numbers}
{}
{Since our model incorporates the strategies of disclosure ($x \neq 0$) and non-disclosure ($x=0$), we compute the reproduction numbers in each case by evaluating the Jacobian of our system at the respective equilibrium points. The results obtained in each case are given below. A detailed calculation for the same is provided in the Appendix.}

\begin{figure*}
     \centering
    \includegraphics[scale=0.39]{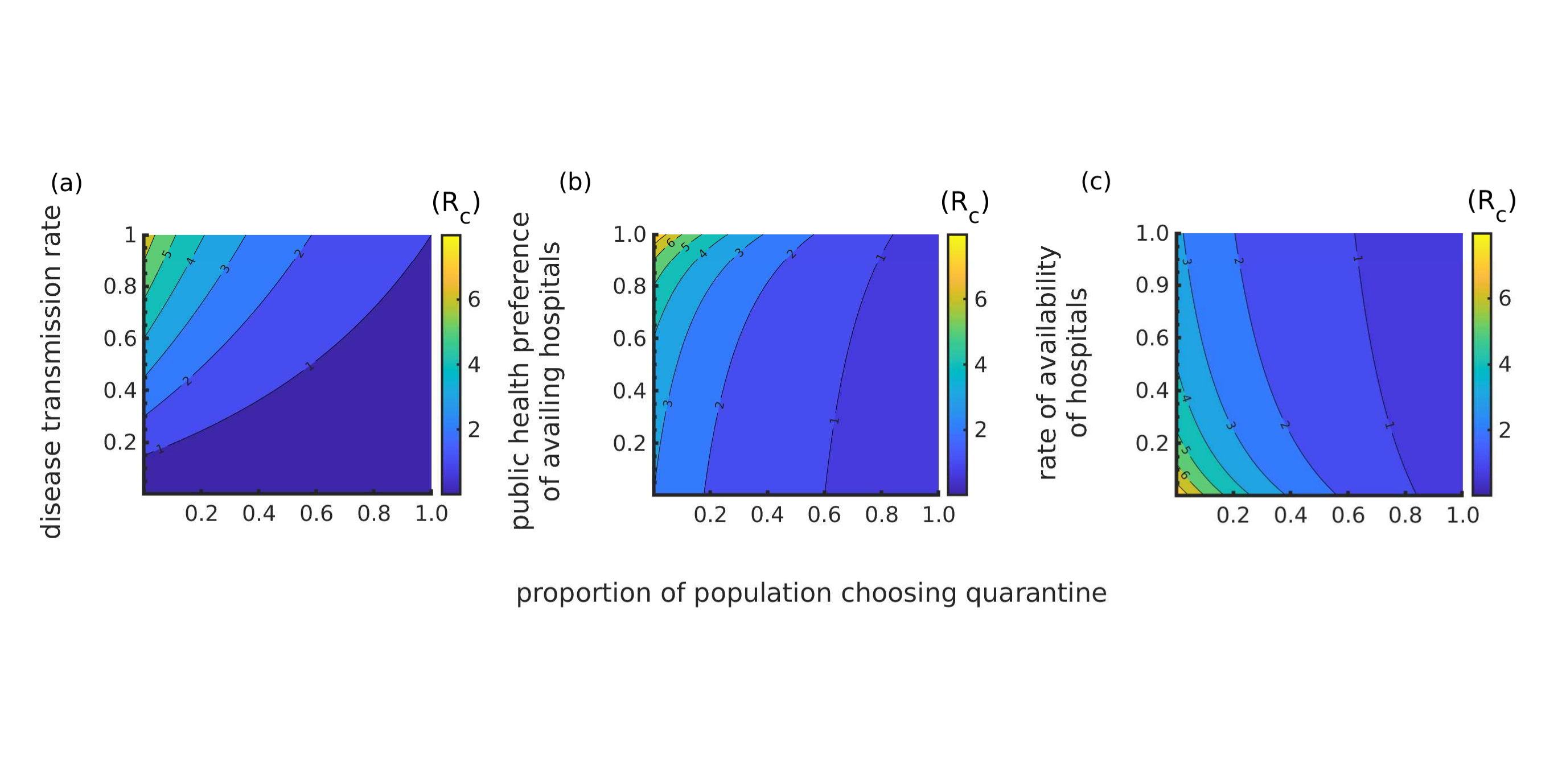}
         \caption{Control reproduction number ($R_c$) of the model for different values of $x: 0 \leq x \leq 1$ on the x-axis and (a) $0 < \beta_s <1$ (b) $0 \leq \eta_q \leq 1$ (c) $0 \leq \alpha \leq 1$ on the y-axis.}
         \label{FIG:R}
\end{figure*}

\begin{itemize}
    \item The basic reproduction number $R_0$ at $x = 0$ is given by:
    \begin{equation}
        R_{0} = \beta_{S}\left[\frac{1}{\beta_e} + \frac{1}{((1-\eta_q)\alpha+\gamma)}\right].
    \end{equation}
    \item Whereas, the control reproduction number $R_c$ when $x \neq 0$ is given by:
    \begin{equation}
        R_{c}(x) = \frac{\beta_s}{(x+\beta_e(1-x))}\left[1 +\frac{\beta_e(1-x)}{((1-\eta_q)\alpha+\gamma)}\right].
    \end{equation}
\end{itemize}
Since the disease-free equilibrium state is stable for $R_c<1$, we compute the threshold proportion of individuals required to disclose their exposure, ensuring that the infection in the population dies out completely. This is given by:
\begin{equation}
    x_c > \frac{((1-\eta_q)\alpha + \gamma)(\beta_s-\beta_e)+\beta_s\beta_e}{\beta_s\beta_e+(1-\beta_e)((1-\eta_q)\alpha+\gamma)}.
\end{equation}
{}
{We have further plotted the control reproduction number $R_c$ as a function of $x$ and other important parameters such as $\beta_s$ (the transmission rate), $\eta_q$ (public health preference in providing hospital beds), and $\alpha$ (rate of hospital beds in the community)} (Figure \ref{FIG:R}). {}{This figure clearly shows that higher values of $x$ reduce the force of infection in the transmission of the disease} (Figure \ref{FIG:R}(a)). {}{Also, higher availability of hospital beds in the community or even relatively low preference in accessing hospitals among individuals with disease will die out immediately with a lower proportion of disclosing} (Figure \ref{FIG:R}(b \& c)). {}{This happens because the infected individuals, once hospitalized, are isolated and reduce the transmission. This indicates that a greater supply of beds in the community might help reduce the transmission of infection. We now simulate the model using several parameters under baseline values to analyze the impact of disclosing infection on the dynamics of disease spread.}

\section{Numerical Simulation}
We simulated the FDE model (\ref{eq:c1}) with the values of parameters described in Table \ref{table1:values_transposed}. {}{We assume a fixed population size of $N = 50100$, with initial population distributions as follows: $E(0) = 0$; $I(0) = 100$; $Q(0) = 0$; $H(0) = 0$; $R(0) = 0$; and $x(0) = 0.15$. Numerical solutions for the FDE system are provided by the Adams-Bashfourth-Moulton scheme, as described in the supplementary section. All computations and simulations were done in Python by developing an entire code of the numerical scheme and applying it to our model. The results were compressed into CSV files and transferred to Matlab for higher-resolution images, and all the corresponding plots were generated.}
\begin{table*}
\begin{center}
\caption{Baseline values (or ranges) of the parameters {}{used for the numerical simulations of the disease model} (\ref{eq:c1})}.\label{table1:values_transposed}
\begin{tabular}{|c|c|c|c|c|c|c|c|c|c|c|c|c|} \hline
 Parameter & $\beta_s$ & $\beta_e$ & $\eta_q$ & $\alpha$ & $\mu_h$ & $\zeta$ & $\gamma$ & $\gamma_h$ & $\kappa$ & $c_d$ & $c_{nd}$ & $p_S$\\ \hline
Values/Ranges & 0.6 & 0.3 & (0-1) & (0-1) & 0.3 & 0.9 & 0.1 & 0.071 & (0-1) & (0-1) & (0-1) & (0-1)\\ \hline
Ref. & \cite{he2020seir} & \cite{niu2020modeling} & & & \cite{ndairou2020mathematical} & \cite{alrabaiah2021comparative} & \cite{chae2020estimation} & \cite{sun2020tracking} & & & & \\ \hline
\end{tabular}
\end{center}
\end{table*}
\subsection{Impact of rate of disease transmission rate ($\beta_s$) and public health intervention ($\eta_q$) on disease prevalence}
\begin{figure*}
     \centering
    \includegraphics[scale=0.38]{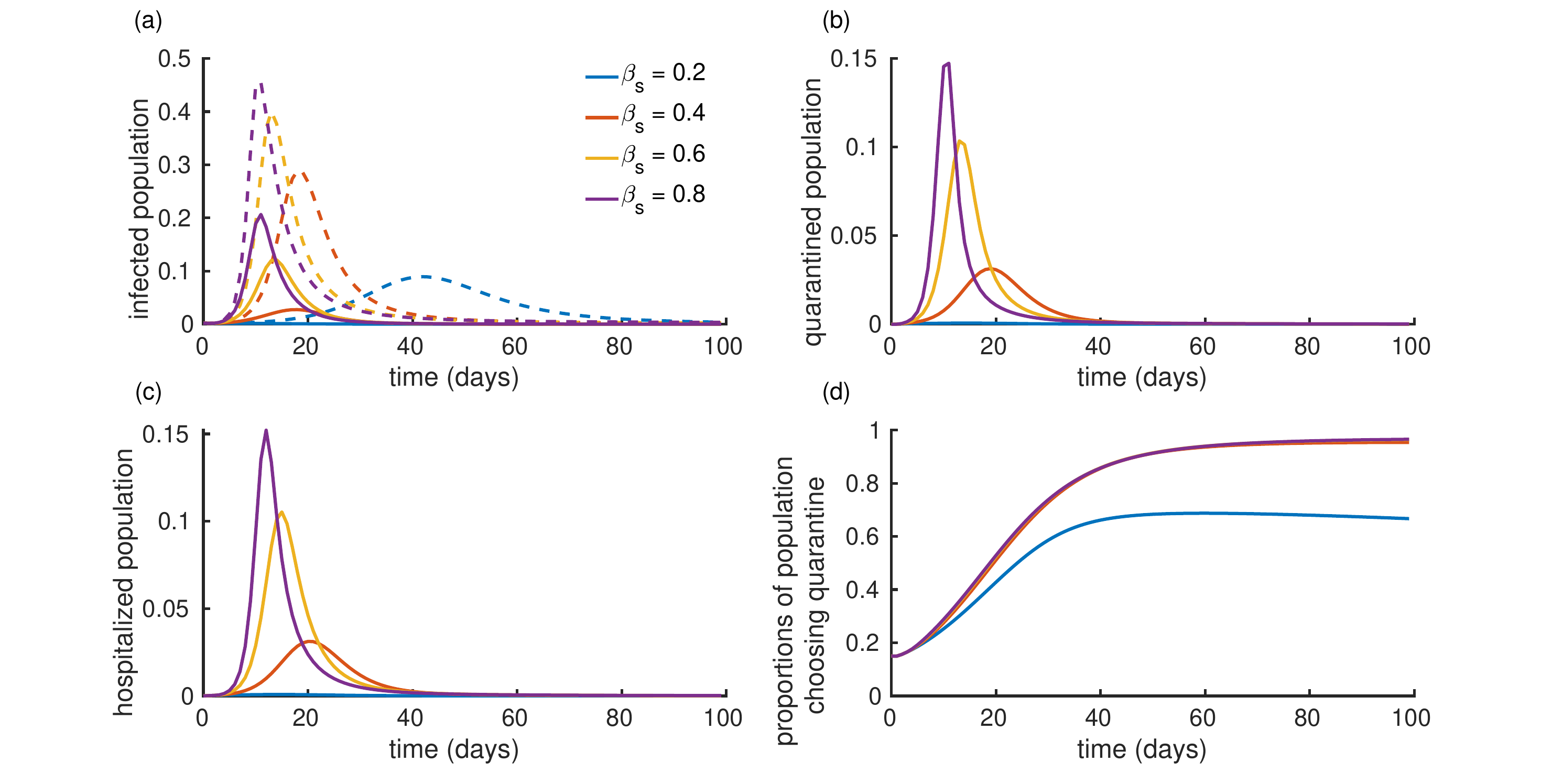}
         \caption{The dynamics of different trajectories of the model {}{for various values of disease transmission rates, $\beta_s$}: (a) infected population, (b) quarantine population, (c) hospitalized, and (d) proportion $x$ of individuals who choose to disclose their infection, with $\kappa = 0.5$, $A = 10$, and $p_s =0.8$. In Figure (a), the dotted line indicates the dynamics of the infected population when {}{no individual chooses to disclose exposure, i.e.,} $x=0$. {}{These comparative trajectories underscore the importance of the decision game on the infection burden.}}
         \label{FIG:R1}
\end{figure*}
Transmission rate of infection ($\beta_{s}$) and disease burden are important factors behind {}{an} individual's decision to adopt quarantine \cite{ashcroft2021quantifying,hossain2020effects}. Figure \ref{FIG:R1} depicts the proportion of individuals in different compartments for sample values of $\beta_{s}$. A relatively small value of $\beta_{s} = 0.2$ would mean a smaller fraction of individuals moving to the exposed compartment daily, subsequently leading to fewer infections, quarantines, and hospitalizations (Figures \ref{FIG:R1} (a),(b),(c)). Whereas, for a larger value of $\beta_{s} = 0.8$, the force of infection increases, causing an increase in the number of individuals in all compartments (Figures \ref{FIG:R1} (a),(b),(c)). There are two important observations in this simulation: The first is the stark difference in the transmission profile of infection in the presence and absence of disclosing options (Figure \ref{FIG:R1} (a), dotted lines). The cumulative difference between these two indicates that disclosing exposure to infected neighbours may help reduce infection transmission by improving hospital bed management or by eventually isolating them from the general population, lowering the risk of infection transmission. The second observation is that the proportion of disclosing individuals at $\beta_s=0.2$ is significantly lower than the cases when {}{$\beta_s$ is higher since more exposed players are present in the latter} (Figure \ref{FIG:R1} (d)). Due to the high population in the infected and quarantined compartment for $\beta_s = 0.8$ (Figure \ref{FIG:R1} (a \& b)), the probability of acquiring a severe disease is high. Consequently, a large hospitalized population can be seen (Figure \ref{FIG:R1} (c)). Hence, the burden of surrounding infections directly influences this decision strategy. As a result, people choose to disclose {}{their exposure}.\\\\
\begin{figure*}
     \centering
    \includegraphics[scale=0.38]{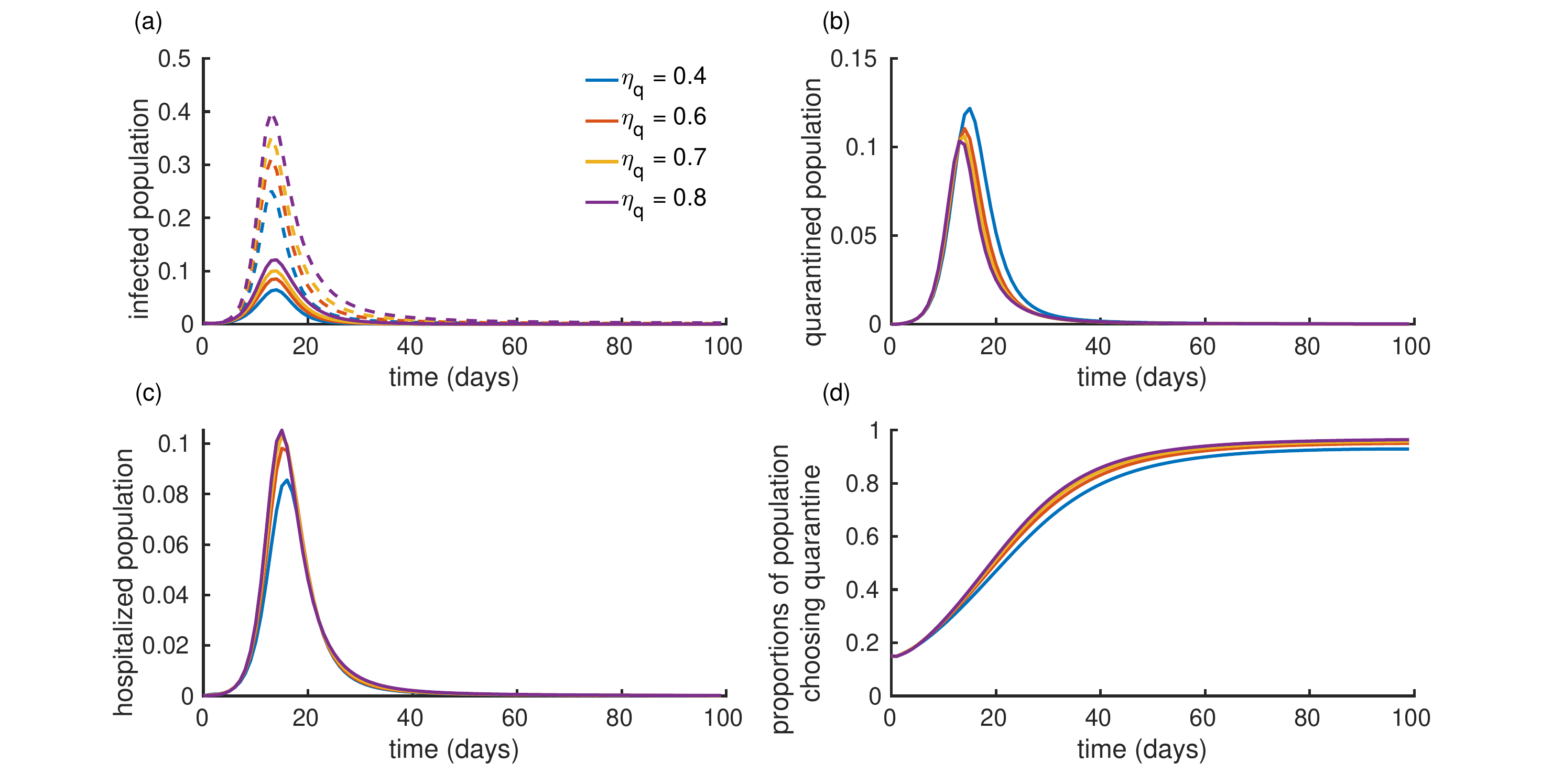}
         \caption{The dynamics of disease component {}{for different values of public health preferences to treatment facilities ($\eta_q$)}: (a) infected population, (b) quarantine population (c) hospitalized, and (d) proportion $x$ of individuals who choose to disclose their infection, with $\kappa = 0.5$ and $p_s =0.8$. In Figure (a), the dotted lines indicate the dynamics of the infected population when {}{no individual chooses to disclose exposure, i.e.,} $x=0$.}
         \label{FIG:R2}
\end{figure*}
One of the public health measures during the COVID-19 epidemic was to provide priority access to hospital facilities for people who voluntarily entered quarantine \cite{duan2020psychological,grimm2020hospital}. Figure \ref{FIG:R2} {}{shows the impact of varying the public health preference factor $\eta_q$ on the disease burden}. A higher value of $\eta_q$ {}{works as an incentive for exposed individuals to choose quarantine since it }indicates a higher probability of getting hospital facilities. On the other hand, a lower $\eta_q$ discourages quarantine {}{as seen in} Figure \ref{FIG:R2} (d). However, a {}{similar trend is not} reflected in the population of the Q compartment (Figure \ref{FIG:R2} (b)) {}{since} individuals are also moving {}{from the Q} to the H compartment {}{at a higher rate for greater values of $\eta_q$}. {}{Also, the number of infections still remain high} (Figure \ref{FIG:R2} (a)), {}{as} higher $\eta_q$ values decrease the chances {}{for an infected individual to get} a hospital facility {}{,since they} do not choose the strategy of disclosure. Of course, with and without disclosing exposure show a similar difference as earlier (Figure \ref{FIG:R1}). {}{This analysis indicates important information for public health policymakers: more hospital facilities for individuals who choose to be quarantined may not reduce the burden of non-disclosing infected individuals in the population to a great extent, causing a larger contribution to the force of infection}.
\begin{figure*}
     \centering
    \includegraphics[scale=0.4]{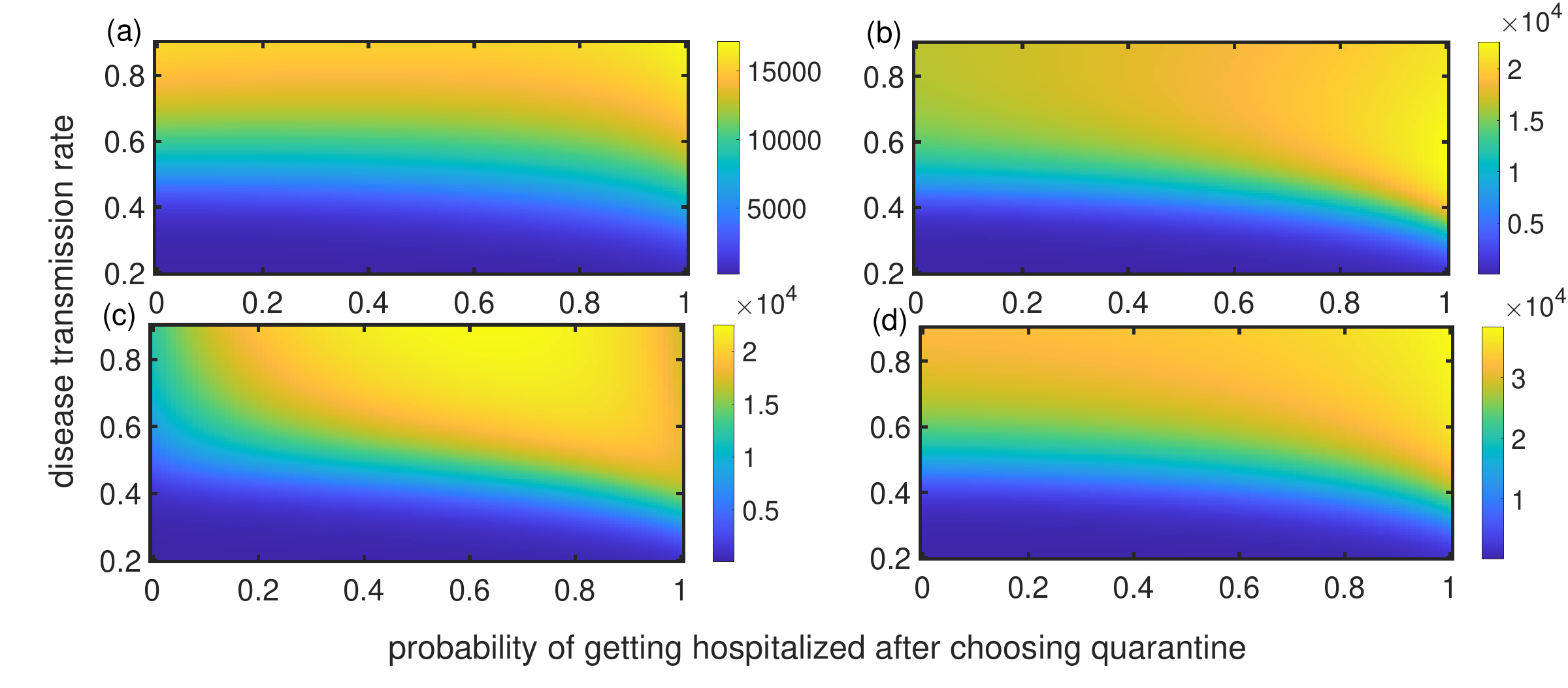}
         \caption{{}{The cumulative new entries of} (a) infection (b) quarantined (c) hospitalized {}{and} (d) {}{population opting to disclose}; as a function of the probability of getting hospitals after choosing quarantine ($\eta_q$) and transmission rate ($\beta_s$).}
         \label{FIG:R11}
\end{figure*}

Despite these results, the disease may be controlled by a combination of disease prevalence, burden, and public health initiatives. The combined influence of these two significant epidemiological factors {}{on the cumulative new entries of each compartment} is summarized in Figure \ref{FIG:R11}. As previously observed, greater $\beta_s$ values increase the disease burden, {}{resulting in a larger entry of individuals in the infected (I) compartment. However, in this case, the cumulative burden remains high irrespective of the values of $\eta_q$, since the new infections do not directly depend on $\eta_q$. This is per our model framework} (Figure \ref{FIG:R1} (a). {}{Since the new hospitalizations depend on $\eta_q$, higher values of $\beta_s$ would imply a greater movement of infected and quarantined individuals into the hospitalized compartment based on public health preferences.} As a consequence of this, hospitals will accommodate the greatest number of patients possible (Figure \ref{FIG:R11}(c)). For $\eta_q \geq 0.899 $ and $\beta_s \geq 0.5456 $, the number of individuals {}{entering} quarantine is very high (Figure \ref{FIG:R11}(d)) because, for the same values, the cumulative exposed population is very large.\\
We {}{also} analyzed the difference between the baseline payoff of disclosure and the new payoff of disclosure without $1-\eta_q$ to figure out the importance of public health measures (Figure \ref{FIG:R11}\, \& \, S1). {}{The} cumulative {}{new} infections with our baseline payoff is $\approx$ $1.5\times10^{4}$ for $\eta_q \geq 0.8$ and $\beta_s \geq 0.7$ (Figure \ref{FIG:R11}(a)); however, without $1-\eta_q$, public health intervention does not affect {}{the decision strategy}, {}{leading to a greater cumulative infection that} varies mainly owing to $\beta_s$ (Figure S1). Public health efforts may help control the disease if there is a high transmission rate. {}{A similar trend in the cumulative quarantines can be seen in }Figures (\ref{FIG:R11}(d), S1(d)) {}{due to the interplay of these factors.}\\
Our research suggests that the public health intervention of providing hospitals for quarantine has a favourable effect on the quarantine and hospital burden if the disease transmission rate is high in the community. Additionally, this can help reduce the prevalence of disease in the population. During a pandemic, public health agencies may implement measures that mitigate the effects of the disease on the general population.
\subsection{Influence of incubation rate ($\beta_e$) and probability of perceived severity ($p_s$) on the disease}
\begin{figure*}
     \centering
    \includegraphics[scale=0.38]{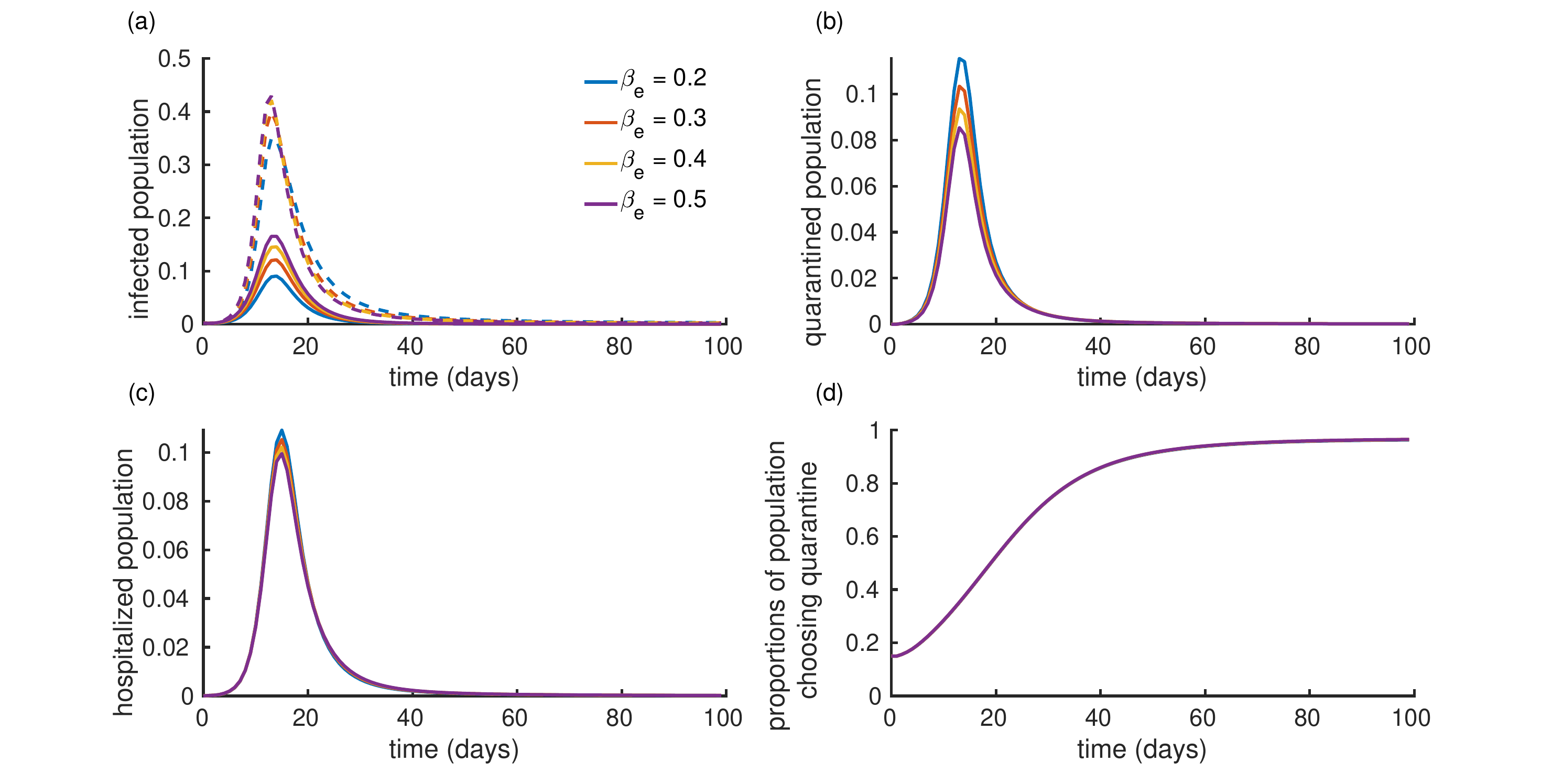}
         \caption{The dynamics of different trajectories of the model: (a) infected population, (b) quarantine population (c) hospitalized, and (d) proportion $x$ of individuals who choose to disclose their infection for different values of incubation rate ($\beta_e$). {}{Here}, $\kappa = 0.5$ and $p_s =0.8$. In Figure (a), the dotted lines indicate the dynamics of infected population {}{when no individual chooses to disclose exposure, i.e., $x=0$}}
         \label{FIG:R3}
\end{figure*}
The rate of developing disease ($\beta_{e}$) after exposure to infection may play a certain role in disease dynamics. In Figure \ref{FIG:R3}(a), a low value of $\beta_{e} = 0.2$ corresponds to a lesser contribution to the infected class, while a higher value of $\beta_{e} = 0.5$ would mean a greater contribution. Different values of $\beta_{e}$ do not affect the decision strategy of disclosing the infection at the onset of the disease {}{since the replicator equation }(\ref{eq:G5}) {}{is independent of $\beta_e$. This is seen in} Figure \ref{FIG:R3}(d)). However, a change in the proportion of quarantined individuals still occurs, as reflected in \ref{FIG:R3}(b). This is because the number of individuals in the exposed compartment is initially constant for a fixed value of $\beta_{s}$. Then, based on non-disclosure, individuals move into the infection compartment at the rate of $\beta_{e}$, leaving a fraction of individuals to move into quarantine by disclosing. So, a lower value of $\beta_{e} = 0.2$ would correspond to an increase in the number of individuals deciding to move into quarantine, as shown by the blue curve in Figure \ref{FIG:R3}(b). A change in the quarantined population also occurs due to the exit of individuals {}{into} the hospitalized and recovered compartments, as given by our model.\\\\
\begin{figure*}

\centering
\includegraphics[scale=0.35]{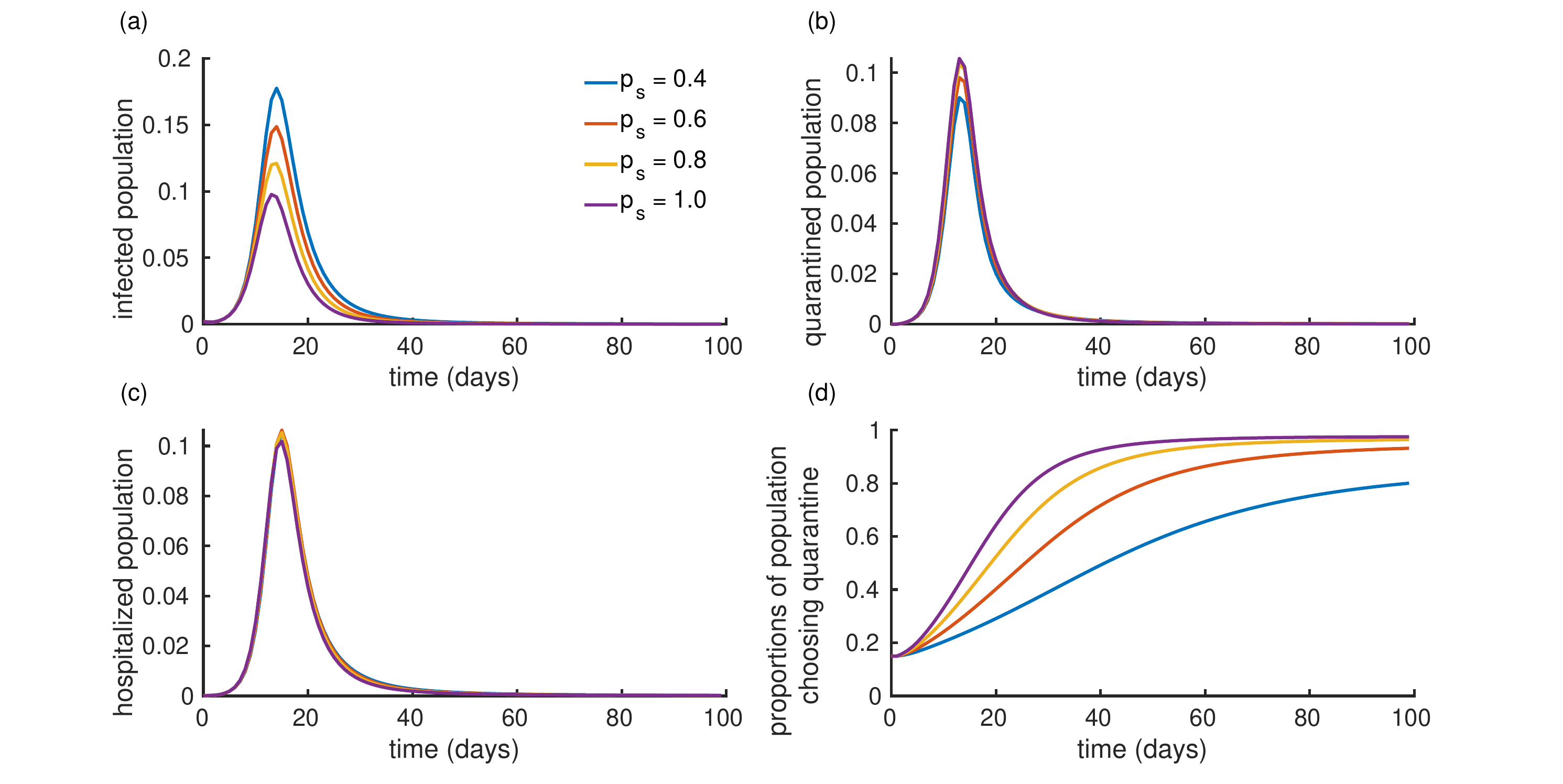}
\caption{The dynamics of different trajectories of the model: (a) infected population, (b) quarantine population, (c) hospitalized, and (d) proportion $x$ of individuals who choose to disclose their infection for different {}{ perceived probabilities of developing severe symptoms} upon infection ($p_s$). {}{Here,} $\kappa = 0.5$.}\label{FIG:R4}
\end{figure*}
Adopting a quarantine approach is heavily influenced by the relative probability of perceived severe symptoms after getting an infection \cite{lei2020comparison,davis2022quarantine}. Figure \ref{FIG:R4} shows the interplay of disease severity and quarantine for different values of $p_s$. When the {}{perceived probability of severity is very high}, i.e., $p_s = 1$, more individuals prefer to adopt quarantine by disclosing their infection (Figure \ref{FIG:R4} (d),(b)). {}{This simultaneously leads to} a lesser proportion of individuals non-disclosing and hence moving into the infected compartment (Figure \ref{FIG:R4} (a)). However, the number of hospitalizations remains unaffected in any scenario of severity, as seen in Figure \ref{FIG:R4} (c) {}{since} the availability of hospital beds relies on $\eta_q$, {}{and has} no direct dependence on this probability. {}{Overall}, these results show that the infection rate influences the number of people in quarantine but does not affect the fraction of people choosing quarantine at any given time. However, if {}{surrounding individuals have a greater probability of developing} a severe illness, the proportion of people who can make decisions shifts dramatically {}{through this perception}. Our findings may prompt public health officials to take action in response to the outbreak by ordering more isolation units to be built.
\subsection{Impact of the perceived probability of severity ($p_s$) with preference by public health authorities to provide treatment facilities ($\eta_q$)}
\begin{figure*}
\centering
\includegraphics[scale=0.35]{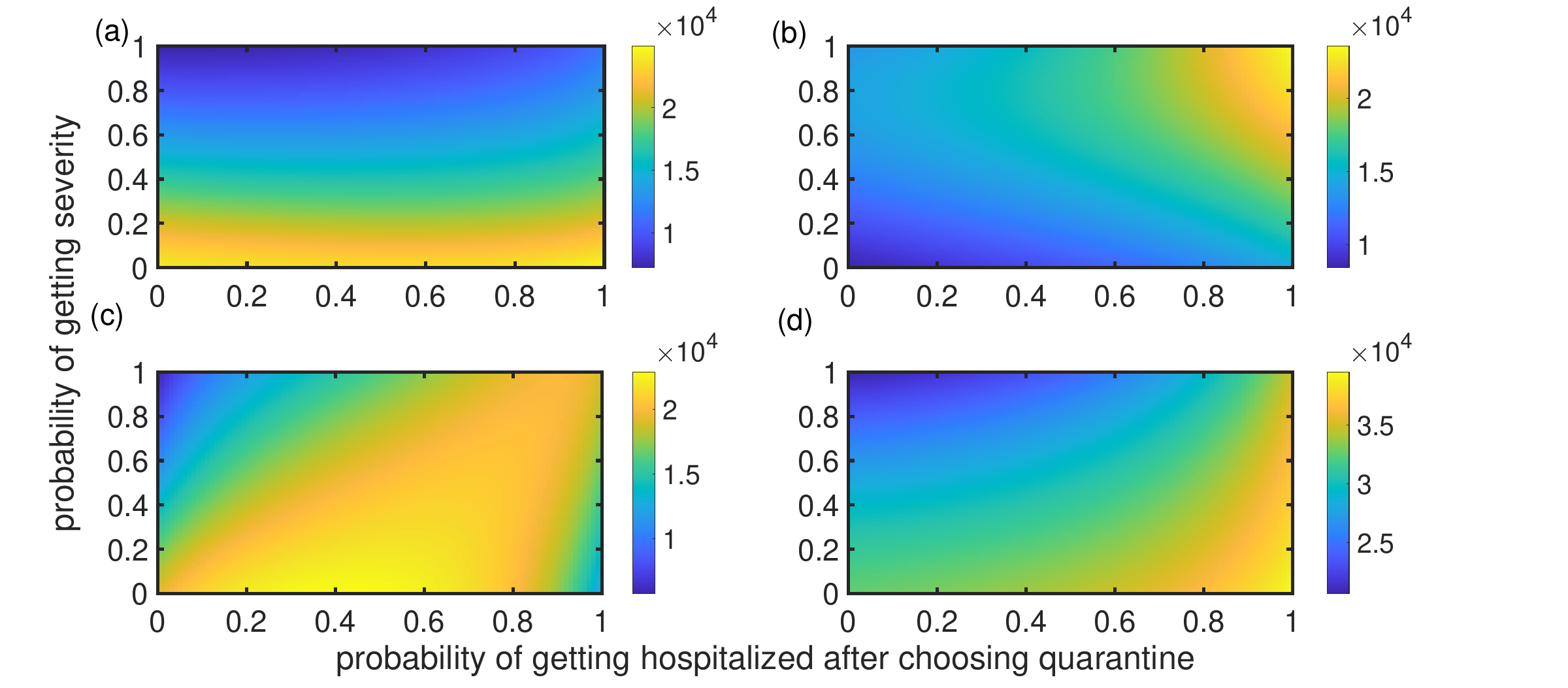}
\caption{{}{The cumulative new entries of} (a) infection (b) quarantined (c) hospitalized {}{and} (d) {}{population opting to disclose}; as a function of preference by public health authorities to provide treatment facility ($\eta_q$) and probability of getting severity ($p_s$).}\label{FIG:R12}
\end{figure*}
 Figure \ref{FIG:R12} summarizes the {}{simultaneous effect of $p_s$ and $\eta_q$ on the cumulative new infections and hospital burden}. For {}{$p_s \leq 0.2$, higher} values of $\eta_q$ indicate that a large proportion of the population will opt for quarantine per day (Figure \ref{FIG:R12} (d)). As the value of $p_s$ rises, people will fairly play with the strategy and make decisions to maximize their payoffs. For $p_s \geq 0.8$ and $\eta_q \geq 0.9$ (Figure \ref{FIG:R12} (d)), a large fraction of the population will engage in the decision-making process, which {}{results in a greater inflow of individuals opting to quarantine} (Figure \ref{FIG:R12}(b). {}{Simultaneously, the inflow of infections reduce due to non-disclosure} (Figure \ref{FIG:R12} (a)). The burden {}{of new hospitalizations also} shrinks, as anticipated, when a significant portion of the public participates in the decision-making process (Figure \ref{FIG:R12} (c)). Therefore, when {}{$p_s \leq 0.2$ and $\eta_q \leq 0.6$} (Figure \ref{FIG:R12} (c)), a significant portion of the population is hospitalized. Concurrently, $p_s$ has a greater influence on the payoff (Figure \ref{FIG:R12} (d)).\\
We also analyzed the difference between the baseline payoff and the new payoff of the disclosure without $1-\eta_q$ to determine the importance of public health measures (Figure \ref{FIG:R12} $\&$ S2). When $\eta_q$ is considered in our model, it mimics how public health interventions reduce the disease burden, whether or not the condition is becoming severe. Based on our research, we hypothesize that the perceived severity of the illness is one of the main drivers of the increase in hospitalizations seen during the pandemic. If public health benefits are greater for the quarantined population, then a large proportion will be involved in switching to disclose exposure. Understanding the true extent of the infection and effective public health messaging can motivate people to opt for quarantine.

\subsection{ Data implementation, parameter estimation and scenario analysis}
\begin{figure*}
\includegraphics[scale=0.35]{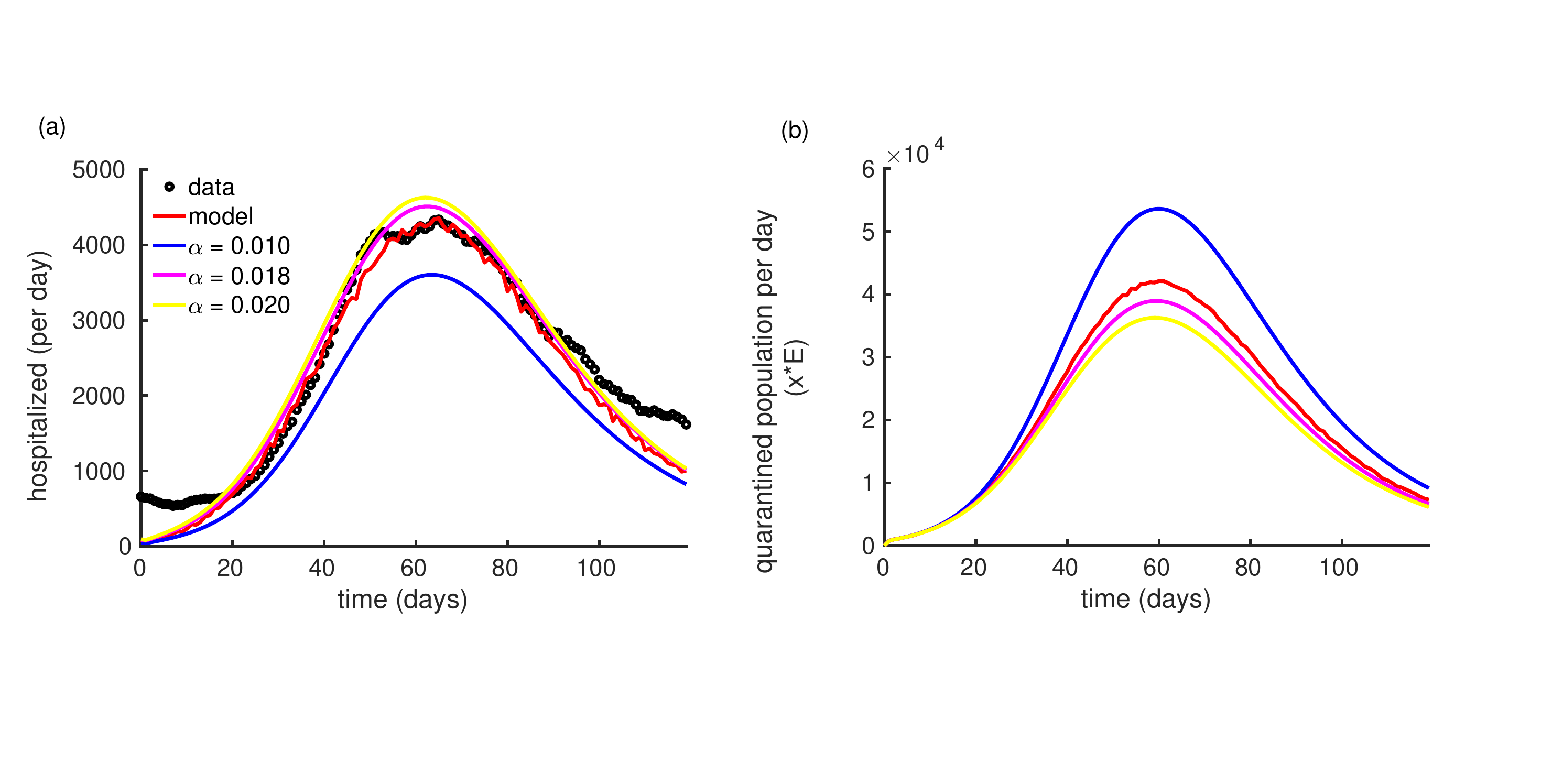}
\caption{ The dynamics of (a) model fit the confirmed new daily {}{hospitalizations }in Chile, South America for the period from April to September 2020; (b) quarantined population per day with estimated parameters {}{given in $\Theta_e$ and corresponding optimized values as shown in Table 3}. {}{The plots also show the scenarios of hospitalizations and quarantines} for different values of {}{hospital admission rate }$\alpha$. }\label{FIG:modelfitting}
\end{figure*}
We used historical hospitalized cases during COVID-19 from Chile, South America to estimate the parameters in our model (\ref{eq:c1}). The total population of Chile was 19.6 million. We considered daily hospitalized cases during the period from April to September 2020. Since the model does not account for vaccination, We chose the time period of the data as the pre-vaccination period. The data set was acquired through the \textit{Our World in Data} \cite{owidcoronavirus}. We estimate mainly {}{freedom of information(FOI) parameters (such as $\eta_q$ which can be accessed in the form of hospital bed availability and treatment priority information via public health portals)} and behavioural parameters (Table \ref{table2:values}) denoted by $\Theta_{e}$:
\begin{equation*}
  \Theta_{e} = \{\beta_{s},\beta_{e},\eta_{q},c_{d},c_{nd},\kappa,A,p_{S},\zeta \}  
\end{equation*}
Other parameters were kept fixed, obtained mainly from the literature (see Table \ref{table3:values}). We used the Least square Approximation (LSA) approach to estimate the parameters.
As depicted in Figure \ref{FIG:modelfitting}(a), our disclosure game model nicely captures the pattern of hospitalized cases, especially the peak timing and the duration of the outbreak. The goodness of fit ($R^{2}$) is approximately 0.95037, indicating our model's relatively strong predictive power. However, the model can't reproduce the pattern in the data at the beginning and end of the outbreak. This is likely because our model ignores comorbidity factors common during COVID-19 hospitalization. According to some statistics, it was observed that 75\% of hospitalized patients with COVID-19 have at least one comorbidity. The most common among these are hypertension, diabetes, cancer, neurodegenerative diseases, cardiovascular diseases, obesity, and kidney diseases \cite{silaghi2023comorbidities}.
\begin{table*}
\begin{center}
\caption{Estimated parameter values {}{(optimized)} in $\Theta_{e}$ {}{for data fitting of daily hospitalizations in Chile.}}\label{table2:values}
\begin{tabular}{cccc} \hline
& & Optimized values\\ Parameters & Description & (unit) \\ \hline
$\beta_s$  & Mean disease transmission rate  & 0.262849 (per day)\\
$\beta_e$  & Mean incubation rate  & 0.97278 (per day)\\
$\eta_q$  & Preference by public health authorities&  0.45296\\
$\zeta$ & Order of fractional derivative &   0.84552\\
$\kappa$ &  Sampling rate in social learning  & 0.016914 (per day)\\
$A$ & Half saturation coefficient & 545.976\\
$c_d$ &  Per unit cost of disclosing infection & 0.09425 \\
$c_{nd}$ &  Per unit cost of non-disclosing infection & 1.5053 \\
$p_S$ &   Probability of developing severity upon infection & 0.918087\\ \hline
\end{tabular}
\end{center}
\end{table*}
The estimated parameter values are given in Table \ref{table2:values}. Along with these parameters, we also estimate the initial condition: $S(0)=19623948.960,E(0) = 0,Q(0)=0,I(0)=8592.1279,H(0)=11.509,R(0)=0, x(0)= 0.35805$. We also calculated the basic reproduction number $R_{0}$ with estimated parameters, which turned out to be 2.6876. Since this value is greater than 1, it signifies an active force of infection, leading to increased hospitalizations, which agrees with our model-fitting result. The detailed estimation method is given in the Supplementary Information (SI). We also plotted the daily disclosing population with the estimated parameters, which estimates around 45000 individuals possibly took part in disclosing their infection and being quarantined during the peak timing of the outbreak (Figure \ref{FIG:modelfitting}(b)). To further examine, we tested the different scenarios of hospital admission rate ($\alpha$) under the estimated parameters. {}{A higher admission rate would imply a lower fraction of individuals left in the quarantine state, as seen in }Figure \ref{FIG:modelfitting}(a-b).
\begin{table*}
\begin{center}
\caption{Fixed parameter values {}{(non-optimized)} in $\Theta_{f}$ {}{for data fitting of daily hospitalizations in Chile.}}\label{table3:values}
\begin{tabular}{cccc} \hline
Parameters & Description & Values (per day) & ref.\\ \hline
$\alpha$  & Rate of hospital admissions & 0.016 & \cite{sen2021closer}\\
$\gamma$  & Recovery rate from infection  & 0.1 & \cite{chae2020estimation} \\
$\gamma_{h}$  & Recovery rate of hospitalized patients & 0.067 & \cite{sun2020tracking}\\
$\mu_{h}$ &   Mortality rate after hospitalization  & 0.011 & \cite{rzymski2023covid}\\ \hline
\end{tabular}
\end{center}
\end{table*}
\section{Discussion}
The COVID-19 pandemic presented a multifaceted challenge to public health systems and communities worldwide \cite{contini2020covid,tunio2021multifaceted,ceci2021development}. Managing the spread of the SARS-CoV-2 virus necessitated a comprehensive approach, with individual actions playing a crucial role in the collective battle against the pandemic \cite{eftekhari2021comprehensive,das2023brief}. Individual actions in early detection and isolation have been consistently highlighted as a critical component of pandemic control \cite{dutta2021local,guner2020covid}. When individuals promptly recognize symptoms, get tested, and self-isolate upon a positive result, they prevent further spread of the virus within their communities \cite{salathe2020covid, thorneloe2022adherence}. However, in most cases, the individual action was voluntary, i.e., public health authorities encourage individuals to disclose the infection and eventually quarantine if they are exposed to other infected people in the community, which turns out to be an individual decision-making initiative because this always incurs some cost. In our research, we have investigated how individual decision-making changes with several aspects of the transmission of infection in society or the availability of hospital beds in the community.

We investigated this decision-making approach by developing a standard SEIR (susceptible-exposed-infected-recovered) model, including hospital and quarantine compartments. We use a fractional derivative approach to represent disease propagation within a population. We have shown that a rise in the rate of disease transmission leads to an increase in the illness burden and the load placed on hospitals. Consequently, more individuals will be involved in the decision-making process. The concomitant impact of public health interventions and disease transmission rates will decrease the population's hospitalization rate, although it increases with the severity of the disease. Our research highlights that quarantine may be a favourable strategy at the individual level in specific epidemiological situations. This may provide insights for public health officials in developing intervention plans to control an epidemic.

There are recent works on game theory modelling social isolation behaviour and its impact on disease control \cite{hossain2020effects, saad2023dynamics}. Our paper takes a distinct approach from these referenced works, although there is significant overlap in motivations and applications. For instance, while Hossein et al. (2021 \cite{khazaei2021disease}) and Saad-Roy et al. (2023 \cite{saad2023dynamics}) presume that susceptible individuals are the players who can choose to adhere to Non-Pharmaceutical Interventions (NPIs), our model assumes that exposed individuals are the players in the game. Consequently, the availability of hospital beds drives behavioural choices in our model, unlike in the other cited research. For the same rationale, the fundamental disparity between our work and theirs lies in how we implement the control of infection transmission. Our approach focuses on reducing infection within the community by decreasing the number of infected individuals. Still, its methods primarily revolve around depleting susceptible individuals from the population to achieve the same goal. Additionally, unlike the referenced works that assume reinfection and demonstrate oscillating dynamics, we have excluded reinfection from our system.

Evolutionary game theory has provided insightful perspectives on the evolution of behaviours during epidemics. Recently, it has applications in diverse emerging fields, encompassing complex networks, biological physics, and cyberphysical systems \cite{yan2024game,roy2023recent,hafezalkotob2023policy,palomo2022agent,ochab2023multiple}. On the other hand, the fractional order technique has been used to model tumour growth, drug transport, immune response, and inflammation dynamics \cite{farman2020analysis,pachauri2023multi,amilo2023mathematical}. We conducted simulations across various fractional order values to compare the dynamical pattern generated by the fractional differential equation model with the ODE model. We compared them with traditional ordinary differential equation (ODE) simulations. Remarkably, as the fractional order (\(\zeta\)) converges towards 1, ODE results exhibit convergence with fractional models (Figure S3). We also found that the fractional-order derivative model fits the hospitalized data better than the ordinary differential equation (ODE) model (Figure S4). Details are given in the Supplementary Information (SI).

While the current research has furnished a valuable framework for understanding the importance of quarantine as a strategy for controlling disease outbreaks, our model is based on simplified assumptions in many aspects. There are scopes for improvement in the analysis. The analysis can be broadened to examine how the periodicity of infection dynamics interacts with the fluctuation of individual choices. Our model simplified the population as uniform in age, not accounting for age-specific behaviours, which is very natural \cite{wittemyer2021differential,addai2023fractional}. To enhance the realism of the quarantine strategy, a heterogeneous population could be integrated \cite{diaz2022heterogeneous}. {}{Furthermore, incorporating time-delays in disease incubation and acquired immunity can assist in capturing the dynamics of disease transmission in a more realistic way}. From the perspective of epidemiological studies, it is also worthwhile to consider the impact of disease severity on infection or even look into infection-disclosing behaviours during the vaccine's availability. {}{Coupling these ideas with the predictive abilities of fractional derivatives will be of tremendous importance in disease outbreaks since better-predicted information can potentially avert severe outcomes.}

Nonetheless, our research provides a forum for discussing individuals' perceptions of disclosing infection at the start of an epidemic. It also explores the causes and consequences of individuals' strategic decision-making in the face of different epidemiological conditions. Individuals always maximize their payoff while ignoring population-level externalities when making such decisions. However, understanding the complex interaction is critical from a public health perspective because it may aid in disease outbreak management and benefit individual health in the community.

\section{Appendix}
\subsection{Equilibria, Reproduction numbers and Stability analysis}
We have the disease-free equilibrium and the disease-endemic equilibrium. Depending on whether individuals opt for disclosing or not, there are mainly two important disease-free equilibria: $E^d_1 = (N,0,0,0,0,0)$ and $E^d_2 = (0,0,0,0,0,1)$. We call $E^d_1$ a purely non-disclosing game and $E^d_2$ a purely disclosing game. We compute the basic reproduction ratio, the control reproduction ratio, and the Jacobian matrix near the disease-free equilibrium $E^d_1$. This gives the optimal fraction of disclosing individuals needed to contain the disease soon.
\subsubsection{Basic reproduction number $R_{0}$ when $x=0$}
In the case when $x=0$, the reproduction number $R_{0}$ is given by the dominant eigenvalue of the next-generation method as follows:\\
The infected classes for this system will be:
\[
^{C}_{0}D^{\zeta}_{t}E(t) = \frac{\beta_s(I+E)S}{N}-\beta_{e}E
\]
\[
^{C}_{0}D^{\zeta}_{t}I(t) = \beta_{e}E-(1-\eta_q)\alpha I-\gamma I
\]
So:
\begin{equation*}
f = 
\begin{bmatrix}
\frac{\beta_s(I+E)S}{N}\\
0 
\end{bmatrix},  
v = 
\begin{bmatrix}
-\beta_{e}E\\
\beta_{e}E-\gamma I-(1-\eta_q)\alpha I
\end{bmatrix}
\end{equation*}
\begin{equation*}
F = 
\begin{bmatrix}
\frac{\beta_sS}{N} & \frac{\beta_sS}{N}\\
0 & 0
\end{bmatrix},
V = 
\begin{bmatrix}
-\beta_{e} & 0\\
\beta_{e} & -(\gamma+(1-\eta_q)\alpha)
\end{bmatrix}
\end{equation*}
\begin{equation*}
V^{-1} = \frac{1}{\beta_e ((1-\eta_q)\alpha+\gamma)}
\begin{bmatrix}
-(\gamma+(1-\eta_q)\alpha) & 0\\
-\beta_{e} & -\beta_{e}
\end{bmatrix}
\end{equation*}
\begin{minipage}[t]{\linewidth}
\small
\begin{equation*}
\implies 
FV^{-1} = \frac{1}{\beta_e ((1-\eta_q)\alpha+\gamma)}
\begin{bmatrix}
\frac{-(\gamma+(1-\eta_q)\alpha)\beta_sS-\beta_e\beta_sS}{N} & \frac{-\beta_e\beta_sS}{N}\\
0 & 0
\end{bmatrix}
\end{equation*}
\end{minipage}
The dominant eigenvalue of the next-generation matrix
[$FV^{-1}$] is:
\begin{equation*}
    R_0(S,N) = \frac{\beta_sS}{\beta_eN} + \frac{\beta_sS}{((1-\eta_q)\alpha+\gamma)N}
\end{equation*}
Evaluating this at the disease-free equilibrium point \\$E^{df}(S^{df},E^{df},I^{df},Q^{df},H^{df}) = (N,0,0,0,0)$  gives the reproduction number $R_{0}:$
\[
R_{0} = \beta_{S}\left[\frac{1}{\beta_e} + \frac{1}{((1-\eta_q)\alpha+\gamma)}\right].
\]
\subsubsection{Control reproduction number $R_{c}$ when $x\neq 0$}
The control reproduction number $R_{c}$ is obtained similarly, computed as follows: \\
The Infected classes are:
\[
^{C}_{0}D^{\zeta}_{t}E(t) = \frac{\beta_s(I+E)S}{N}-xE-\beta_{e}(1-x)E
\]
\[
^{C}_{0}D^{\zeta}_{t}I(t) = \beta_{e}(1-x)E-(1-\eta_q)\alpha I-\gamma I
\]
So:
\begin{equation*}
f = 
\begin{bmatrix}
\frac{\beta_s(I+E)S}{N}\\
0 
\end{bmatrix},
v = 
\begin{bmatrix}
-xE-\beta_{e}(1-x)E\\
\beta_{e}(1-x)E-\gamma I-(1-\eta_q)\alpha I
\end{bmatrix}
\end{equation*}
\begin{equation*}
F = 
\begin{bmatrix}
\frac{\beta_{s}S}{N} & \frac{\beta_sS}{N}\\
0 & 0
\end{bmatrix},
V = 
\begin{bmatrix}
-x-\beta_{e}(1-x) & 0\\
\beta_{e}(1-x) & -((1-\eta_q)\alpha+\gamma)
\end{bmatrix}
\end{equation*}
$\implies V^{-1} = \frac{1}{(x+\beta_e(1-x))((1-\eta_q)\alpha+\gamma)} \times$\[
\begin{bmatrix}
-(\gamma+(1-\eta_q)\alpha) & 0\\
-\beta_{e}(1-x) & -(x+\beta_{e}(1-x))
\end{bmatrix}
\]\\
$\implies FV^{-1} =\frac{1}{(x+\beta_e(1-x))((1-\eta_q)\alpha+\gamma)}\times$
\[\begin{bmatrix}
\frac{-(\beta_sS)*((1-\eta_q)\alpha+\gamma)-\beta_e(1-x)\beta_sS}{N} & \frac{-(x+\beta_e(1-x))\beta_sS}{N} \\
0 & 0
\end{bmatrix}
\]
Thus, the dominant eigenvalue of the next-generation matrix
[$FV^{-1}$] is:
\begin{equation*}
    R_{c}(x) =\\ \frac{\beta_sS}{(x+\beta_e(1-x))N}+\frac{\beta_e\beta_s(1-x)S}{(x+\beta_e(1-x))((1-\eta_q)\alpha+\gamma)N}.
\end{equation*}
Evaluating this at the disease-free equilibrium point $E_1^d$, gives the control-reproduction number $R_{c}$ as a function of $x$:
\[
R_{c}(x) = \frac{\beta_s}{(x+\beta_e(1-x))}\left[1 +\frac{\beta_e(1-x)}{((1-\eta_q)\alpha+\gamma)}\right].
\]
The disease-free equilibrium is stable if $R_{c}<1$, which implies
\begin{equation}
x_c > \frac{((1-\eta_q)\alpha + \gamma)(\beta_s-\beta_e)+\beta_s\beta_e}{\beta_s\beta_e+(1-\beta_e)((1-\eta_q)\alpha+\gamma)}
\end{equation}
This indicates that it requires at least an $x_{c}$ proportion of individuals to disclose their exposure to infection for the infection in the population to die out immediately.  \\
We also compute the Jacobian matrix of the system at $E^d_1$:\\
\noindent
\begin{minipage}{\linewidth}
  \centering
  \resizebox{\columnwidth}{!}{
    $\displaystyle
    J^{d}=
    \begin{bmatrix}
        \frac{-\beta_s(I+E)}{N} & \frac{-\beta_sS}{N} & 0 & \frac{-\beta_s S}{N} & 0 & 0\\
\frac{\beta_s(I+E)}{N} & \frac{\beta_sS}{N} - x - \beta_{e}(1-x)  & 0 &\frac{\beta_sS}{N}& 0&-E+\beta_{e}E\\
0 & x &-\eta_q \alpha - \gamma & 0 & 0 & E\\
0 & \beta_{e}(1-x) & 0  & -(1-\eta_{q})\alpha - \gamma& 0&-\beta_e E\\

0 & 0 &\eta_{q}\alpha & (1-\eta_{q})\alpha & -\gamma_{h}-\mu_h&0\\
0 & 0 & 0 & 0&-\frac{\kappa a c_{nd}p_s x(1-x)}{(a+H)^2} & (1-2x)\kappa \Delta E

    \end{bmatrix}
    $
  }
\end{minipage}

Substituting the values of the equilibrium component into the Jacobian matrix, we get:\\
\begin{minipage}{\linewidth}
  \centering
  \resizebox{\columnwidth}{!}{
    $\displaystyle
    J^{d}=
    \begin{bmatrix}
0 & -\beta_s & 0 & -\beta_s &0&0\\
0&\beta_{s} -\beta_e & 0 & \beta_s & 0 & 0\\
0 & 0 & -\eta_{q}\alpha-\gamma & 0&0&0\\
0&\beta_{e} & 0&-(1-\eta_{q})\alpha - \gamma  & 0 &0\\
0&0 &\eta_{q}\alpha& (1-\eta_{q})\alpha & -\gamma_{h}-\mu_h&0 \\
 0 & 0 & 0 & 0 &0&-\kappa c_{d} (1-\eta_{q})

 \end{bmatrix}
   $
  }
\end{minipage}

It has the following eigenvalues:
\begin{align*}
\lambda_{1} = 0,\;
\lambda_{2} =  -\eta_{q}\alpha - \gamma,\;
\lambda_{3} = -\gamma_{h}-\mu_h,\;\\
\lambda_{4} = -kc_d(1-\eta_{q})\;.
\end{align*}
Other two eigenvalues are represented by the following quadratic equation in $\lambda$:
\begin{align*}
    \lambda^2 & - \lambda\left[(\beta_s-\beta_e)-((1-\eta_q)\alpha+\gamma)\right]+ \\&\left[\beta_s\beta_e - (\beta_s-\beta_e)((1-\eta_q)\alpha+\gamma)\right] = 0.
\end{align*}
The following sufficient criterion ensures that these eigenvalues are also negative:
\begin{equation*}
    \beta_s - \beta_e < min\{\sqrt{\beta_s \beta_e},(1-\eta_q)\alpha+\gamma\}
\end{equation*}
This, however, indicates the stability of the disease-free equilibrium $E^d_1$.

\section*{CRediT authorship contribution statement}
\textbf{Pranav Verma:} Coding, Methodology, Formal analysis, Writing – original draft. \textbf{Viney Kumar:} Coding, Methodology, Formal analysis, Writing – original draft. \textbf{Samit Bhattacharyya:} Conceptualization, Methodology, Formal analysis, Writing – review \& editing, Supervision.
\section*{Data availability}
The COVID-19 case data used in this manuscript is open-access data.
\section*{Acknowledgments}
The authors thank all reviewers for their insightful comments and suggestions for improving the presentation of the content in the manuscript. This work was initiated under the Shiv Nadar Institution of Eminence (SNIoE)'s Opportunity for Undergraduate Research (OUR) program. Viney Kumar thanks the Council of Scientific and Industrial Research (FILE NUMBER 09/1128(12030)/2021-EMR-I), India, for its financial support. All authors acknowledge SNIoE DST-FIST-funded computational lab support. \cite{bauch2005imitation, bhattacharyya2011wait, bauch2012evolutionary}
\section*{Conflict of interest}
The authors declare that they have no known competing financial interests or personal relationships that could have
appeared to influence the work reported in this paper.

\end{document}